\documentclass[preprint, numberedappendix]{emulateapj}
\usepackage{natbib}

\newcommand{\etal}{et al.\ }
\newcommand{\fun}{{\rm cm^{-2}\, s^{-1}}}

\begin{document}

\title{Unresolved Unidentified Source Contribution to the Gamma-ray
  Background} 

\author{V. Pavlidou\altaffilmark{1,2}, 
        J. M. Siegal-Gaskins\altaffilmark{1,3}, 
        B. D. Fields\altaffilmark{4},  
        A. V. Olinto\altaffilmark{1,2,5}, 
        C. Brown\altaffilmark{3} %etc.
}

\altaffiltext{1}{Kavli Institute of Cosmological Physics, 
The University of Chicago, Chicago, IL 60637}
\altaffiltext{2}{Enrico Fermi Institute, The University of Chicago, 
Chicago, IL 606137}
\altaffiltext{3}{Department of Physics,
The University of Chicago, Chicago, IL 60637}
\altaffiltext{4}{Department of Astronomy, 
University of Illinois at Urbana-Champaign, Urbana, IL 61801}
\altaffiltext{5}
{Laboratoire Astroparticule et Cosmologie (APC), 
Universit\'e Paris7/CNRS, 10 rue A. Domonet L. Duquet,
75205 Paris, Cedex13, France}

\begin{abstract}

The large majority of EGRET point sources remain without an identified 
low-energy counterpart, and a large fraction of these sources are most
likely extragalactic. Whatever the nature of the extragalactic 
EGRET unidentified sources, faint unresolved objects of the same 
class must have a contribution to the diffuse extragalactic 
gamma-ray background (EGRB).  Understanding this component 
of the EGRB, along with other guaranteed contributions 
from known sources, is essential if we are to use 
this emission to constrain exotic 
high-energy physics.
Here, we follow an empirical approach to 
estimate whether a potential contribution of unidentified sources to the EGRB is likely to be important, and we find 
that it is. Additionally, we show how upcoming GLAST observations of 
EGRET unidentified sources, as well as of their fainter counterparts, can be 
combined with GLAST observations of the Galactic and extragalactic 
diffuse backgrounds to shed light on the nature of the EGRET 
unidentified sources even without any positional association of such 
sources with low-energy counterparts. 


\end{abstract}
\maketitle

\section{Introduction}
\label{intro}

The origin of the isotropic diffuse emission  (Sreekumar et al.\ 1998) in energies between 
$100 {\, \rm MeV}$ and $20 {\, \rm GeV}$, detected by the  
{\it Energetic Gamma-Ray Experiment Telescope} (EGRET) aboard 
the {\it Compton Gamma-Ray Observatory}, remains one 
of the great unknowns of GeV-energy astrophysics. There are two
major questions that still remain unanswered. 1.\ How much of the
diffuse emission detected by EGRET is, in fact, extragalactic, and what is the spectrum of this extragalactic background? And 2.\ what fraction of the
extragalactic emission can be attributed to each of the observationally
established classes of gamma-ray emitters? Despite the associated 
uncertainties, these two issues are critical in any attempt to use
gamma-ray observations to constrain exotic high-energy physics and
yet-undetected classes of theorized gamma-ray emitters. 

To answer the first question, a good understanding of the
Galactic diffuse emission in the EGRET energy range is required. 
In order to obtain the intensity and spectrum of the extragalactic emission 
from the EGRET sky maps, the Galactic emission needs to be modeled and
subtracted. This is made complicated by the discrepancy between 
the observed Milky Way spectrum in energies of $\gtrsim 1 {\rm GeV}$ 
and theoretical expectations (Hunter et al.\ 1997). The observed spectrum is 
more shallow than model predictions based on 
the local demodulated cosmic ray spectrum. This deviation is known as
the ``GeV excess'', and although various explanations have been
proposed to account for part or all of the discrepancy, its origin
remains a matter of debate (e.g., Pohl et al.\ 1997; B\"{u}sching
  et al.\ 2001; Strong et al.\ 2004b; 
Kamae et al.\ 2005; de Boer et al.\ 2006; Strong 2006; Stecker \etal 2007). 
As a result, determinations of the
gamma-ray background using different Galactic emission models yield
very different answers, both in intensity and in spectrum, despite being
based on the same set of observations (e.g., Sreekumar \etal 1998; 
 Strong et al.\ 2004a). 

Attempts to answer the second question have been plagued by
uncertainties in the cosmic density and evolution
of the two established classes of extragalactic gamma-ray 
emitters: normal galaxies and blazars. Our observational knowledge of 
the gamma-ray properties of normal galaxies is very limited, as the sample of
normal galaxies which have been observed in gamma rays consists of 
only the Milky Way and a marginal detection of the Large Magellanic Cloud
(Sreekumar et al.\ 1992; Hunter et al.\ 1997; Hartman et al.\ 1999). For
this reason, the accuracy of theoretical estimates of the contribution 
of normal galaxies to the gamma-ray background  
is unavoidably at the order-of-magnitude level (e.g., Lichti et
al.\ 1978; Pavlidou \&
Fields 2002). But even in the case
of blazars, which are by far the most numerous and best studied class
of identified gamma-ray emitters, estimates of their contribution 
to the gamma-ray background vary from a few percent to $100\%$
of the background originally reported by the EGRET team 
(e.g., Padovani \etal 1993; Stecker \& Salamon 1996a; 
Kazanas \& Perlman 1997; Mukherjee \&
Chiang 1999; M\"{ucke} \& Pohl 2000; Narumoto \& Totani 2006;
Dermer 2007). 

The issue is further complicated by the existence of 171 sources
which, at the time of publication of the 3rd EGRET catalog (hereafter
3EG; Hartman \etal 1999), 
had not been positively or potentially 
associated with a lower-energy counterpart. 
These sources are collectively known as the unidentified EGRET
sources, and they are more numerous than any established group of gamma-ray
emitters. The distribution of these sources on the sky is such that a
Galactic feature can be clearly distinguished - however a large number
of sources are located away from the Galactic plane and the Galactic
center\footnote{Note however that the presence of sources at high latitudes does not, in itself, constitute proof that these objects are extragalactic (see e.g.\ the Gould Belt discussion in Gehrels et al.\ 2000).}. No more than a handful of sources
can be associated with the Milky Way halo if the Milky Way is not
many times brighter in gamma-rays than similar galaxies such as M31
(Siegal-Gaskins \etal 2007). Hence, it is
almost certain that the EGRET unidentified sources include a
significant extragalactic component. 
Although the nature of these extragalactic sources
remains unknown, it is reasonable to 
believe that there is a large number of fainter, unresolved objects 
of the same class, which are guaranteed to have {\em some} contribution to 
the extragalactic gamma-ray background (EGRB). If these sources represent
yet unidentified members of some known class of gamma-ray emitters
(e.g.\ blazars), then excluding them from any calculation of 
the contribution of the parent class to the diffuse background
would lead to a significantly underestimated result 
due to an incorrect normalization of the bright-end of the
gamma-ray luminosity function. 
If they represent an unknown class of gamma-ray emitters,
then the contribution of their
unresolved counterparts to the diffuse emission would significantly
limit the diffuse flux left to be attributed to known classes, exotic
processes, and  truly diffuse emission. 

Hence, some contribution of unresolved unidentified sources to the 
EGRB is certain. It is therefore clear 
that until we either  answer the question of the nature of
unidentified sources or derive some strong constraint 
indicating that a possible contribution of such
unresolved objects to the EGRB would indeed be minor, we cannot 
hope to fully understand the origin of the EGRB.

Detailed predictions for 
the level of the unidentified source contribution to the EGRB
involve important uncertainties: since no 
low-energy counterparts have been identified, we have no estimates
of distance, and therefore no estimates of the gamma-ray 
luminosities of these sources. As a result,
very few constraints can be placed on their cosmic distribution 
and evolution. However, very simple estimates can offer some guidance on whether ignoring this EGRB component may be a safe assumption to make. 

For example, we can use the number of unidentified sources, the minimum flux resolvable by EGRET, and the observed intensity to the extragalactic gamma-ray background to place rough limits on the distance scales associated with resolved and unresolved unidentified sources so that unresolved sources do not overproduce the background. 
A population of unbeamed, non-evolving, single-luminosity sources uniformly distributed in Euclidian space are resolvable out to a distance $D$ by an instrument of number flux sensitivity $F_{\rm min}$. The relation between $D$, $F_{\rm min}$, and number luminosity $L$ in this case is simply $L=4\pi D^2F_{\rm min}$. If the instrument detects $N$ such sources, their number density $n_{\rm source}$ can be estimated to be $n_{\rm source} = 3N/4\pi D^3$. If the same distribution of sources continues out to a distance $d>D$, the isotropic intensity (photons per unit area per unit time per unit solid angle) from the {\em unresolved} members of this population will be $I_{\rm unres} = \int_D^d dI_{\rm shell}$, where $dI_{\rm shell} = (1/4\pi) (n_{\rm source}4\pi r^2 dr) L/(4\pi r^2)$ is the contribution from sources within a spherical shell located at a distance $r$ from the observer. Substituting our results for $n_{\rm source}$ and $L$ above, and performing the integral, we obtain
\begin{equation}
I_{\rm unres} =  3NF_{\rm min}(d-D)/4\pi D\,.
\end{equation}
If we require that the unresolved emission from this population does not exceed the EGRB observed by EGRET ($I_{\rm unres} \leq I_{\rm EGRB}$), we obtain  $I_{\rm EGRB} \geq 3NF_{\rm min}(d-D)/4\pi D$. Substituting $N\sim 100$ for the population of extragalactic unidentified sources (see discussion in \S \ref{samples}), $F_{\rm min} \sim 10^{-7} {\rm \, ph \, cm^{2} \, s^{-1}}$ for the sensitivity of EGRET, and $I_{\rm EGRB} \sim 10^{-5}  {\rm \, ph \, cm^{2} \, s^{-1} \, sr^{-1}}$ (Sreekumar et al. 1998), we obtain $d \lesssim 6D$. This result implies that the largest distance out to which such a distribution of objects persists cannot be larger than a few times the distance out to which these objects are currently resolved, since in any other case these sources would overproduce the EGRB. It is therefore conceivable that the unidentified source contribution to the EGRB is significant, if not dominant. 

In this work, we approach the problem from a purely empirical
point of view. Instead of attempting to {\em predict} the level of a
diffuse component due to unresolved objects of the same class as
unidentified EGRET sources, we try to assess whether there are any
empirical indications that this component is, in fact, minor. 
We construct samples of unidentified sources 
which, based on their sky distribution,  are
likely to consist mostly of extragalactic objects. 
Under the assumption that the majority of these
 sources can be treated as members of a single 
class of gamma-ray emitters, we seek to answer
the following three questions:

(1) Is it likely that unresolved objects of the same class could have
    a significant contribution to the EGRB at least in some energy range?  

(2) How would the collective spectrum of their emission compare to the
    measured spectrum of the EGRB 
    deduced from EGRET observations? 

(3) How are GLAST observations expected to improve our understanding of
the nature of unidentified sources, based on the insight gained from
our analysis?

This paper is structured as follows. In \S \ref{samples} we discuss
the samples of resolved unidentified sources used in our analysis. Our
formalism for constructing the collective emission spectrum of
unresolved unidentified sources is presented in \S
\ref{formalism}. Inputs from EGRET data used in our analysis are
described in \S \ref{inputs}. In \S \ref{results} we describe our
results, and in \S \ref{sec:GLAST} we discuss how these results are
expected to improve once GLAST observations become available. Finally,
we summarize our conclusions in \S \ref{sec:disc}.

\section{Source Samples}\label{samples}

\begin{figure}
\begin{center}
%\resizebox{!}{2.8in}
%{
\plotone{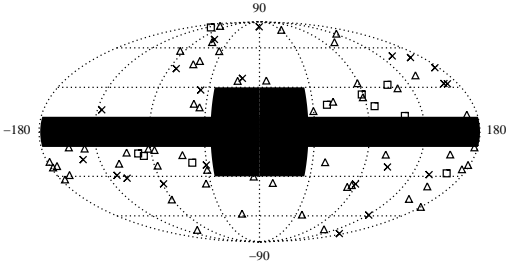}%}
  \caption{\label{mask}
Source exclusion mask that has been applied in our analysis. Sources
with coordinates within the shaded regions have been excluded from our
analysis. The locations of sources in our samples are also
indicated. $\times$: possible 3EG blazars (included in Sample 1 only); 
$\square$: originally unidentified in 3EG, with later suggested
associations (included in
Samples 1 and 2); $\triangle$: sources included in all three samples
(unidentified in 3EG, no counterpart suggested to this day).}
\end{center}
\end{figure}

The observational input constraining our calculations will be the 
fluxes of the unidentified sources and
the spectral index of each source. 
The fluxes we use for our analysis are the P1234 photon fluxes
quoted in 3EG. These fluxes
 were determined from the combined maps from all 
viewing periods during which any particular object was within the
EGRET field of view. The spectral
index $\alpha$ of each object is also taken from 3EG, and was 
derived through a fit assuming that the
photon flux of each object obeys $F_{1234}=F_0(E/E_0)^{-\alpha}$ where
$E$ is the photon energy.

We seek to constrain a possible contribution to the EGRB
from objects of the same class
as unidentified sources. 
For this reason, we would like to base our calculations on 
 a {\em resolved} source sample which is as
representative as possible of the population of {\em extragalactic}
unidentified EGRET sources. However, the population of sources which,
at the time of publication of 3EG, did not have
confident associations with known astrophysical objects, clearly
exhibits a Galactic feature, with many of the sources concentrated
around the Galactic plane and at small Galactic longitudes (close to
the Galactic center). Thus, the original population of EGRET
unidentified sources is not an appropriate sample as it includes many
sources which are most likely Galactic. 

Since these sources are unidentified, there is no counterpart information which could be used to distinguish between Galactic and
extragalactic objects. Additionally, inferences based on their
spectral indices and variability are also very hard to make due to the
large error bars associated with measurements of these quantities
(see Siegal-Gaskins et al.\ 2007). Finally, both Galactic and extragalactic
identified objects exhibit a large range of luminosities, so that inferences
about the location of unidentified sources based on their fluxes are not possible either. The only relatively robust indication concerning the location of these sources is their position on the sky. A Galactic
population associated with the disk and bulge will most likely be
concentrated around low Galactic latitudes and small Galactic
longitudes. On the other hand, a halo population, which may appear
more uniformly distributed on the sky, cannot have too many members
among the resolved EGRET unidentified sources,
based on the gamma-ray properties of galaxies similar to the Milky Way
(Siegal-Gaskins et al.\ 2007). Thus, excluding objects around the
Galactic plane and the Galactic center will most likely minimize any
Galactic contamination of the sample of unidentified sources. 

For this reason, we have excluded from our studies all sources with
Galactic latitude $|b|<10^\circ$ for Galactic longitudes
$|l|>40^\circ$ and Galactic latitude $|b|<30^\circ$ for Galactic
longitudes $|l|\leq 40^\circ$. The source exclusion map we have applied
is plotted in Fig.\ \ref{mask}, and is identical to the mask applied by
Sreekumar et al.\ (1998) in their selection of EGRET diffuse emission data to
use for the derivation of the EGRB. The solid angle associated with this mask is $\Omega_{\rm
  mask} \approx \pi$, leaving $\approx 3\pi$ of allowed solid angle. 
 For this reason, and under the assumption that
an extragalactic population will be uniformly distributed on the sky,
we have renormalized all derived flux distributions by a factor of
$4/3$. 

Additionally, we have excluded from all samples objects  J0516+2320,
J0852-1216, J1424+3734, and J2241-6736 because no P1234 flux was quoted for them in 3EG, and objects J0824-4610, J0827-4247,
J0828-4954, J0841-4356, J0848-4429, and J0859-4257 because they were
marked as possible or likely artifacts in 3EG.

We have applied further
selection criteria which result in three distinct samples 
of resolved unidentified sources. The use of different samples will
allow us to test the sensitivity of our results to the details of the
included members. 

Sample 1 includes all sources 
which were unidentified at the time 3EG was
published {\em and} all sources which were characterized as ``possible
blazars'' (identification code ``a'' in 3EG). These 
latter objects are routinely not included in studies of the blazar
luminosity function, so it is reasonable to examine the potential
contribution of this class to the EGRB  by
including it in our analysis. Sample 1 includes 84 sources. 

Sample 2 only includes sources which were unidentified at the time 3EG
 was published. In this case, we do not consider
``possible blazars'' to belong to this sample. Sample 2 includes 63
sources, and is a subset of Sample 1. 

Sample 3 is the most conservative of our samples, in that it contains
the smallest number of objects. In this case, not only have we
excluded all  ``probable blazars'', but also any object for which even
a tentative or  potential identification has been suggested since the
publication of 3EG\footnote{Sample compiled and
 actively maintained  
by C. Brown and available online
at http://GeVsky.org. 
We stress that suggested associations with lower-energy
counterparts included in this compilation are simply results reported
in recent publications; no effort was made to evaluate the
significance and validity of these counterparts by a single uniform standard.}. 
 Sample 3 includes 53 sources, and is a subset of Sample
2. Sources which are present in Sample 2 but not in sample 3 are
the following: J0010+7309, J0724-4713, J1212+2304, J1825+2854,
J1835+5918, J0329+2149, J0725-5140, J1249-8330, J1621+8203,
and J1959+6342. The locations of all the sources in our samples are shown
in Fig.\ \ref{mask}.

\section{Formalism}
\label{formalism}

A quantity which, under the assumption of isotropy, fully describes
the EGRB is the
differential photon intensity $I_E$ (photons per unit
area-time-energy-solid angle). In this section, we calculate $I_E$ for a population of unresolved sources.
We make the following assumptions for the population of unresolved gamma-ray sources we consider.
\begin{enumerate}
\item Each of the sources has an energy spectrum which, in the EGRET and GLAST energy
ranges, can be well described as a single power law of spectral index
$\alpha$, and a total photon flux $F_8$ at energies $E>100 {\rm \, MeV}$ (where $F_8$ is the photon flux in units of $10^{-8} {\rm\, photons \,   cm^{-2} \, s^{-1}}$). 
\item The cumulative flux distribution, $N(>F_8)$ (number of
sources with flux greater than $F_8$ versus $F_8$), of this population 
can be adequately described, between fluxes $F_{8,\rm
  min}$ and $F_{8,\rm max}$, as a power law, 
\begin{equation}\label{fit1}
N(>F_8) = CF_8^{-\kappa}\,.
\end{equation}
\item The probability distribution of spectral indices
of the population $p(\alpha)$ is independent of source luminosity and does not evolve with redshift. 
\end{enumerate}

The differential flux of each object of this population, $F_E$, is related to $F_8$ through 
\begin{equation}
F_8 = \int_{E_0}^{\infty} \!\!\!\!F_EdE = F_{E_0}\int_{E_0}^{\infty} \!\!\!\!\!\!dE\left(\frac{E}{E_0}\right)^{-\alpha} = \frac{F_{E_0}E_0}{(\alpha-1)}
\,,
\end{equation}
provided $\alpha>1$.
Each  unresolved source with a flux $F_8$ 
at energies $>E_0$, has a contribution 
$I_{E,1}(F_8, \alpha)$ to the diffuse emission  which is given by 
\begin{eqnarray}
I_{E,1}(F_8, \alpha) &=& 
\frac{F_E}{4\pi} = \frac{F_{E_0}}{4\pi}
\left(\frac{E}{E_0}\right)^{-\alpha}\nonumber \\ &=& 
\frac{(\alpha-1) F_8}{4\pi(E_0/{\rm GeV})}
\left(\frac{E}{E_0}\right)^{-\alpha}\!\!\!\!
 \fun \, {\rm sr^{-1}} {\rm \, GeV^{-1}}\,,\nonumber \\
\end{eqnarray}
where $\alpha$ is the spectral index of the source, and $E_0 = 100
{\rm  \, MeV}$ in the case of 3EG fluxes. 
The $1/4\pi$ normalization factor comes from assuming an
isotropic distribution of sources, the collective emission of which is
uniformly distributed over the celestial sphere. 

The differential flux distribution of the population is
\begin{equation}
\left|\frac{dN}{dF_8}\right| = \kappa C F_8^{-\kappa-1}\,.
\end{equation} 

The collective diffuse emission due to this population of 
unresolved sources with fluxes between $F_{\rm 8,min}$ and $F_{\rm 8,
  max}$ will be:
\begin{eqnarray}\label{contr}
I_E (E) &=& \int _{F_{\rm 8,min}}^{F_{\rm 8,max}}\!\!\!\! \!\!\!
dF_8 \int_{\alpha=-\infty}^{\infty} \!\!\!\! \!\!\! d\alpha \,\,
\left|\frac{dN}{dF_8}\right| p(\alpha) I_{E,1}(F_8, \alpha) \nonumber \\
&=&  I_0 
\left[\frac{\kappa CF_8^{-\kappa +1}}{-\kappa +1}
\right]_{F_{\rm 8, min}}^{F_{\rm 8, max}} \!\!\!\!\!\!\times
\int_{\alpha = -\infty}^{\infty} \!\!\!\!\!\!d\alpha \, p(\alpha)
\left(\frac{E}{E_0}\right)^{-\alpha} \!\!\!\!\!\! (\alpha-1)\,, 
\nonumber \\
\end{eqnarray}
where $ I_0 = 10^{-7} \rm cm^{-2} s^{-1} sr^{-1} GeV ^{-1} / (4\pi)$. 
Physically, the upper-limit flux $F_{\rm 8,max}$ represents the
sensitivity limit of the telescope (objects with flux higher than this
are resolved and do not contribute to the background), while the 
lower-limit flux $F_{\rm 8,min}$ represents the flux below which the
approximation of Eq.\ (\ref{fit1}) for the cumulative flux distribution
breaks down. The features of the spectral shape of this emission in the simple case when $p(\alpha)$ is a Gaussian are discussed in  Appendix \ref{shape}.

\section{Inputs from EGRET data}\label{inputs}

Equation (\ref{contr}) can provide a first simple estimate for
the possible contribution of unresolved sources of the same class as
extragalactic unidentified EGRET sources to the isotropic diffuse
gamma-ray background. The required inputs are the cumulative flux
function of Eq.\ (\ref{fit1}), and the spectral index distribution of
the source class, $p(\alpha)$. In this section, we derive these inputs from
the three source samples discussed in \S \ref{samples}. Possible source variability has not been accounted for in this analysis. A discussion of possible effects of source variability, as well as our reasoning for not considering them for the purposes of our study, are presented in Appendix \ref{Var}. 

\subsection{The cumulative flux distribution}\label{cfd}

\begin{figure}
\begin{center}
%\resizebox{3.3in}{!}
%{
\plotone{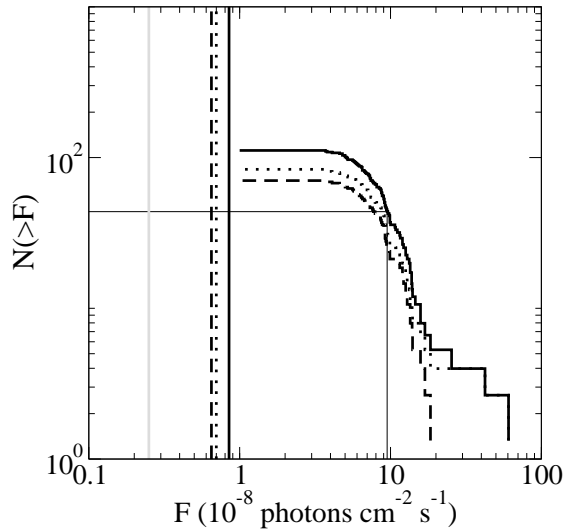}%}
  \caption{\label{lognlogs}
Cumulative flux distribution of the resolved unidentified
sources. Solid line: sample 1; dotted line: sample 2; dashed line:
sample 3. The thin solid lines represent the location of the median value of sample 1 for fluxes below $6\times 10^{-8} {\rm \, photons \, cm^{-2} s^{-1}}$. 
The vertical lines correspond to the lowest flux to which we
can extrapolate a power-law fit to the cumulative flux function of each sample in the flux range  $F_{8, \rm max} < F_8 < F_{8,  N=5}$, before the cumulative emission from unresolved point sources exceeds the measurement of the EGRB at low
energies. The light grey vertical line shows the lowest flux to which we can perform the same extrapolation if the flux function of sample 1 combined with identified blazars outside our mask is used instead.}
\end{center}
\end{figure}

We use data from the resolved objects in our three samples
to construct the cumulative flux distribution of EGRET unidentified
sources.  The cumulative flux distributions of the three samples are shown
in Fig.\ (\ref{lognlogs}). Due to non-uniformities in EGRET exposure, the lowest resolvable flux is not a hard-set number, but rather the efficiency in resolving low-flux objects decreases gradually as the flux decreases, which is one of the major factors in the slow turnover of the cumulative distribution function. For this reason, we only use data within the range $F_{\rm 8, max} < F_8
< F_{8, N=5}$  to obtain a fit to the cumulative flux distribution. 
We take $F_{\rm 8, max} = 6$, and
assume that all sources with fluxes above this limit have been 
resolved by EGRET. We define $F_{8, N=5}$ to be the flux for which, in each
sample, 5 objects with fluxes higher than or equal
to it have been resolved. The best-fit values of $C$ and $\kappa$ for
each sample  are shown in Table~\ref{sometable}. As we can see in
Fig.\ (\ref{lognlogs}) as well as from the parameter values in Table
\ref{sometable}, the slopes of the cumulative flux distribution in the
regime where most objects are expected to have been resolved by EGRET
are fully consistent with each other. This in turn implies that the
data are consistent with the hypothesis that most members of all samples 
have been drawn from a single population of extragalactic
emitters. 

However, we note that the flux distribution slope, close to $\sim$2.5
in all cases, is steeper than the slope that would be expected from a
single-luminosity population of sources with a uniform distribution in
a flat cosmology (which is equal to $1.5$, e.g.\ Dermer 2007), and is also steeper than
the slope of the flux distribution of the confidently identified
blazar population. This result may reflect the cosmological
distribution and evolution properties of the population of
extragalactic unidentified sources. However, it may also
originate in a selection effect in the identification of gamma-ray
sources: the positional identification of brighter sources is more
frequent as more photons are detected from these sources which allows
for a more accurate pinpointing of their location. In
addition, multiwavelength campaigns for the identification of
gamma-ray sources naturally target the brightest objects first. For
these reasons, the high-flux end of the flux distribution may be
preferentially depleted, which would lead to an apparent steepening of the
distribution of the sources that remain unidentified. Finally, the steepness of the slope could be an effect of viewing a distribution with real curvature in a very small dynamical range in flux. A simple example quantifying such a possible effect is the following: 
if we were to treat all sources in Sample 1 and all confidently
identified blazars which are located outside our exclusion mask as a single population, then the cumulative flux distribution power-law fit 
in the flux range $6 <F_8 < F_{8, N=5}$ would have a slope $\kappa =
1.63  \pm 0.03$, much closer to $1.5$. 

Finally, we should add a cautionary note on fits to the cumulative, rather than differential, flux distribution function, such as the one presented here. Our fits purposfully do not account for uncertainties in each bin since, from a statistical point of view, such uncertainties are almost perfectly correlated (each bin contains the same data as its adjacent ones plus/minus one data point). On the other hand, the poor dynamical range in flux and the dependence of uncertainties of the {\em differential} flux function on the size of the bin make constructing and fitting the differential flux function problematic as well. For these reasons, it is important to not over-interpret these results, which should be viewed as broad, order-of-magnitude assessments of the behavior of the flux distribution, rather than robust statistical evaluations and strict constraints.

\subsection{The spectral index distribution}

The second important observational input in our calculation 
is the distribution of spectral indices of unidentified
objects $p(\alpha)$. In the limit that the spectral index of sources is independent of source redshift and luminosity, the spectral index distribution uniquely determines the spectral shape of the collective emission from the unresolved population. The spectral shape of the unresolved emission provides, in turn, an additional tool to assess the possibility that unidentified sources constitute a dominant contribution to the EGRB through {\em shape comparison} between the observed and the predicted spectra of diffuse extragalactic emission. 

In our analysis we adopt the assumption that the
spectral index distribution of unresolved unidentified sources is the
same as that of the resolved unidentified sources. The latter 
can be deduced from measurements of the spectral index $\alpha$
for sources in each one of our samples, following the method presented in Venters \& Pavlidou (2007). The details of this calculation, properly accounting for measurement uncertainties in individual spectral indices of resolved sources, are presented in Appendix \ref{spindex}. 

\begin{table}
\begin{center}
\caption{Fit parameters of the cumulative flux distribution \label{sometable}}
\begin{tabular}{|l|lll|}
\hline
& Sample 1 $\,\,\,\,\,$ 
$\,\,\,\,\,$
& Sample 2 $\,\,\,\,\,$ $\,\,\,\,\,$& 
Sample 3 $\,\,\,\,\,$ $\,\,\,\,\,$\\
\hline
$\ln C$ & $8.79 \pm 0.26$ & $8.93 \pm 0.19$ & $9.18 \pm 0.17$ \\
$\kappa$ & $2.48 \pm 0.12$ & $2.45 \pm 0.09$ & $2.44 \pm 0.07$ \\
\hline 
\end{tabular}
\end{center}
\end{table}

\section{Results}\label{results}
\begin{figure}
\begin{center}
%\resizebox{3.2in}{!}
%{
\plotone{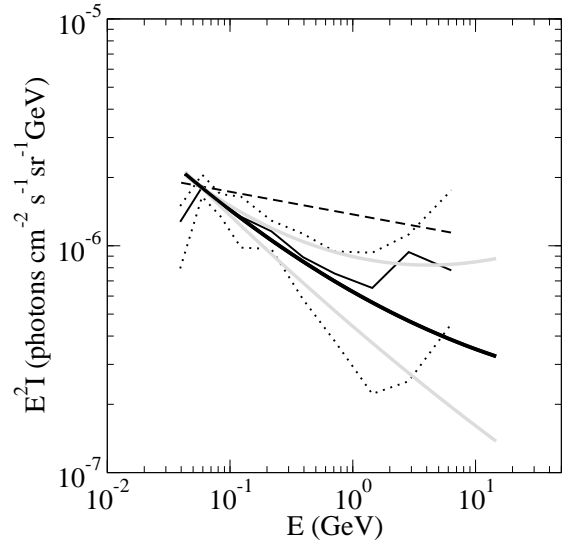}
%}
  \caption{\label{ress}
Dashed line: Sreekumar \etal (1998) determination of the
EGRB. Solid line: Strong \etal (2004a) determination of the EGRB (best
guess). Dotted lines: Strong \etal (2004a) determination of the EGRB
(systematics-based limits). Thick solid line: collective spectrum {\em shape} of unresolved unidentified sources using maximum-likelihood intrinsic spectral index distribution (ISID) parameters for Sample 1 (this work). The amplitude of the spectrum does not constitute a prediction; rather, it has been maximized so that it does not exceed the observed EGRB in any energy bin above 100 MeV.  Thick grey lines: $1\sigma$ uncertainty in the spectral shape of unresolved unidentified source emission for Sample 1, entering through the statistical uncertainty in the ISID parameters.}
\end{center}
\end{figure}

As is immediately obvious from Eq.\ (\ref{contr}), 
the shape and the overall normalization $I_0$ of the cumulative emission
from unresolved unidentified sources are decoupled under our
assumptions. The shape of the spectrum depends on the spectral index
distribution, while the normalization, given a fit to the flux
distribution, depends only on $F_{\rm 8,min}$ (the value of the flux
where the extrapolated power law breaks). 

Assuming that, close
to the EGRET flux limit, the flux distribution does not evolve 
drastically, an extrapolation of the measured flux function to lower
fluxes can be considered representative of its behavior in the 
low-flux regime. This assumption is less likely to hold as the  
limiting flux to which we are extrapolating becomes lower: the flux 
distribution {\em will} eventually  exhibit a break due to
cosmological effects and/or luminosity evolution. 
The question is
how far in the low-flux regime our extrapolation must continue 
before we get a significant contribution of unresolved unidentified 
sources to the gamma-ray background. 
If the answer is ``not very far'', then our extrapolation may indeed be reasonable over such a small range, and unresolved, unidentified sources are likely make up a considerable fraction of the EGRB. On the other hand, if we need to extrapolate the flux
distribution down to fluxes very low compared to the resolved flux
range, then it is quite unlikely that our extrapolation is valid
throughout the flux regime in which we use it.  In such a case it is 
doubtful that the actual flux distribution  
of the unresolved unidentified sources is
such that unidentified sources make up a significant portion of
the EGRB. 

We use observations of the EGRB to
determine by what value of $F_{\rm 8,min}$ the extrapolated flux
distribution power-law {\em must} break, so that the background is
not overpredicted at any energy range. To ensure that we are placing
the most conservative constraint (i.e.\ that we allow extrapolation to
the lowest fluxes possible), we will use the most generous
observational determination of the EGRB,
which is that of Sreekumar et al.\ (1998). The
requirement that this background is not exceeded for any energy above
100 MeV returns the limiting breaking fluxes indicated by the three
vertical lines in Fig.\ \ref{lognlogs}. 
Our results then indicate that, for any of our three samples, 
we would only need to extrapolate the cumulative flux distribution for
at most an order of magnitude below
the lower limit of the resolved flux range to have
unresolved unidentified sources comprise most of the EGRB, at least at
low energies. This is not an extreme extrapolation, and therefore a
significant  contribution by the ``unidentified'' class to the diffuse
background is likely.  

In order to interpret this result correctly, we should assess the qualitative and quantitative uncertainties associated with it. First of all, a power-law extrapolation of the flux distribution to lower, unresolved flux values even for only one order of magnitude in flux is by no means self-evident, although it is the simplest and most straight-forward assumption. For example, the Dermer (2007) fits to the flux distribution of resolved blazars (using a model based on the cosmological evolution of black hole jet sources) increase less steeply than a power law even very close to the EGRET sensitivity limit. It is conceivable that unresolved members of the extragalactic unidentified source class could have a comparable cosmic evolution, and a similar behavior in the flux function. In fact, our tight constraints on the power-law extrapolation of the flux function may be taken to imply just that: the cosmological evolution of these sources {\em has} to be such that the flux function deviates from a power-law form very fast, to ensure than unresolved members of this class do not overproduce the EGRB. 

Second, statistical and systematic uncertainties in the power-law fits to the cumulative flux function have to be taken into account. In our case, the systematic uncertainties associated with the sample selection and the small dynamical range in flux over which the flux function is sampled dominate over the statistical uncertainties in the power-law slope derived in the selected flux range, shown in Table \ref{sometable}. The most extreme case is the one where unidentified sources are considered as a unified sample with identified blazars outside the mask. In this case, as discussed in \S \ref{cfd}, the flux function slope is considerably shallower, allowing for a more extended extrapolation to low fluxes before the EGRB is overproduced. The lowest flux limit allowed for the extrapolation in this case is shown with the grey line in Fig.\ \ref{lognlogs}, and is a moderate factor $\sim 3$ below the limits of the unidentified source samples when considered alone. 

A comparison between the {\em shape} of the cumulative emission
spectrum of unresolved unidentified sources and the EGRET EGRB is
shown in Fig.\ \ref{ress}, 
The dashed line shows a single power-law fit
to the Sreekumar et al.\ (1998) determination of the EGRB. The thin solid line 
is the more recent redetermination of the gamma-ray background by  
Strong et al.\ (2004a), in which they used 
a more detailed model of the Milky Way diffuse emission to 
subtract the Galactic component from the EGRET diffuse sky map. 
The dotted lines are  the systematic uncertainties in the EGRB 
determination of Strong et al.\ (2004a), 
entering through their model of the Galaxy. Our 
calculation of the spectrum of the unresolved unidentified source
component is shown with the thick solid line. The maximum-likelihood
parameters from Sample 1 were used for  $p(\alpha)$ in this
calculation, although, as becomes clear from Fig.\ \ref{cont} in the Appendix, the
spectral shapes resulting from all three samples are consistent with
each other. 
At low energies, where the systematics are low, the spectrum of the unidentified
component is in excellent agreement with the
EGRB observational spectrum of Strong et al.\ (2004a). At higher energies, 
where the systematics are
large, the unidentified component spectrum is largely within
systematics except at very high energies. 
If unidentified sources are indeed a dominant contribution at
relatively low energies, then this result may be perceived as a
tantalizing hint that at the
highest energies of the EGRET range a new type of contribution,
(e.g., from high-energy peaked BL Lacs, or from annihilating dark
matter) may become important at $\sim 1{\rm \, GeV}$. 
 
It should be noted that the amplitude of the cumulative spectrum plotted in Fig.\ \ref{ress} does not constitute a prediction; rather, it is the maximum possible emission allowed from unresolved unidentified sources so that the observed EGRB is not exceeded {\em in any energy bin above 100 MeV}. In Fig. \ref{ress} we also overplot, with the thick grey lines, the $1\sigma$ uncertainty in the spectral shape of the unresolved emission from unidentified sources, entering through uncertainties in the ISID parameters. 

\section{Prospects for GLAST}\label{sec:GLAST}

The launch of GLAST in 2008 will provide significant new
insight into the nature of unidentified sources and their possible
contribution to the EGRB. 
The ideal solution to the unidentified source puzzle would be, of
course, the direct positional association of all unidentified sources
with undisputed low-energy counterparts.  This would then allow us to
build more confident models for the unresolved members of these
classes of objects. However, such an outcome is unlikely, as 
the large number of possible counterparts and the large number of
sources which we expect GLAST will be able to resolve make
multi-wavelength campaigns for every single source impractical. 

However, there is another definitive test that GLAST will be able to
perform which does not require confident identification of each
source to provide information about the likely nature
of unidentified sources as a population. With the increased 
flux sensitivity of GLAST, many more objects of the same class will be
resolved. 
If these objects are mostly extragalactic, as assumed here, and if
they comprise a significant fraction of the EGRET EGRB, as suggested
by our analysis, then there will be an
associated decrease of the EGRB  from its
EGRET levels, equal 
to the all-sky-averaged intensity of the newly resolved objects. 
The flux sensitivity of GLAST is expected to be about 50 times
better than that of EGRET\footnote
{http://www-glast.stanford.edu/mission.html}. 
Therefore GLAST will be able to probe the
flux distribution of unidentified sources down to fluxes close to
$F_{\rm8, min}$ and definitively test our empirical estimate. 

If, on the other hand,  even high-latitude unidentified sources
 are mostly Galactic (which would imply that the Milky Way  may be unusually
 bright in gamma-rays), then there
will be an associated reduction of the Milky Way diffuse emission
rather than of the isotropic background  
(for a discussion on a possible contribution of a large number of 
Galactic point sources to the Galactic diffuse emission and the 
role of such sources in
explaining at least in part the origin of the GeV excess, see 
Strong 2006). 

GLAST will not only represent a leap in instrument sensitivity, but also in energy range, as it will allow access to the yet-unexplored range between a few tens and a few hundreds of GeV. Sources with fluxes high enough that measurements of their spectra can be extended in this range will have an additional indicator of their location (Galactic vs extragalactic): if they are located at cosmic distances, a feature generated by attenuation of gamma-rays on the extragalactic background light may be detectable; on the other hand, if they are located within the Milky Way, no such feature should be present. 

In addition, GLAST observations will improve the inputs used in our
calculation. Spectral indices, at least for the brighter sources, will
be measured with much greater accuracy. This will improve our
constraints on $p(\alpha)$ and the shape of the cumulative emission
from unidentified sources. Additionally, spectral discrimination
between different classes of unidentified emitters may become
possible. Finally, the dynamical range over which fluxes will be
measured will increase significantly, and any biases in the
calculation of the slope of the flux distribution of high-latitude
unidentified sources will become less pronounced. 

\section{Discussion and Conclusions}\label{sec:disc}

In this work, we have used a purely empirical model to explore
 the possibility that unresolved
gamma-ray sources of the same class as unidentified EGRET sources have
an appreciable contribution to the EGRB. 
We have argued that some unidentified source contribution 
to the gamma-ray background is guaranteed. We have additionally found
that 
(1) if most high-latitude unidentified sources are assumed to be extragalactic, 
     a one order of magnitude extension of the cumulative flux distribution 
     to lower energies without breaks implies a significant contribution to 
     the EGRB, at least at
   the lower part of the EGRET energy range; and 
(2) the spectrum of the cumulative emission of such unresolved sources
    would be very consistent with the observational
    determination of Strong et al.\ (2004a)  of the EGRET EGRB
 within systematics. 

We emphasize that the purpose of this study is not to estimate, even at the order-of-magintude level, the diffuse flux expected from extragalactic unresolved unidentified sources, but rather to place constraints on the flux distribution of these objects under the constraint that the measured background should not be exceeded in any energy interval. Our treatment is therefore different in purpose and spirit from most past work aiming to estimate the level of the contribution of different populations to the extragalactic gamma-ray background. 
Such work has been traditionally of two types. The first involves population models built from some understanding of the physics of the sources (such as, e.g., the models of M\"{u}cke \& Pohl 2000 and Dermer 2007 in the case of blazars; Miniati 2003, Keshet et al.\ 2003, Blasi, Gabici, \& Brunetti 2007 in the case of clusters of galaxies; Lichti, Bignami \& Paul 1978 and Pavlidou \& Fields 2002 in the case of normal galaxies). The second involves models of the population luminosity function based on our knowledge of the source population from other wavelengths and normalized to fit EGRET data (such studies require a sample of detected, identified members, and are therefore applicable to blazars only, e.g. Chiang et al.\ 1995, Stecker \& Salamon 1996, Mukherjee \& Chiang 1999, Narumoto \& Totani 2006). 
In our case, lacking any knowledge of the physics of sources as well as of even the bright end of the luminosity function (since in absence of identifications no redshift and hence no luminosity can be derived for any of the sources), we have reversed the problem. Instead of using some assumed cosmic evolution for the sources to derive the expected level of contribution to the EGRB, as in all of the investigations mentioned above, we have used the tightest possible constraint on the allowable EGRB contribution of these sources (the observed EGRET background) to constrain the flux distribution of the unresolved sources. Our conclusions for the overall expected unresolved unidentified source intensity come from the observation that our constraints on the flux distribution are indeed very tight; hence, the contribution of the unidentified sources to the EGRB is likely to be high. 

Our analysis suggests that  any model of the EGRB
 would be incomplete without some
treatment of the unidentified source contribution. 
The results of our empirical model therefore motivate the pursuit of
specific population models for the unidentified sources. 
Although such models involve  a more restrictive set of
assumptions and increased uncertainties, they can provide more 
concrete predictions for the luminosity function of unresolved
objects. Once a luminosity function model is assumed, and under the assumption that unresolved members of this class do indeed contribute most of the extragalactic diffuse emission, additional estimates can be made regarding the distance scales associated with this population.  The dynamical range in fluxes of the {\em resolved} members of the population, as can be seen in  Fig. \ref{lognlogs}, is smaller than an order of magnitude. For a single-luminosity population, this corresponds to a factor of 3 dynamical range in distance, which increases or decreases if the typical source luminosity increases or decreases with increasing redshift respectively. Such arguments can constrain not only the distance scales, but also the number density, intrinsic brightness, and evolution of the resolved and unresolved objects, if in a model-dependent fashion. 
Additional constraints and predictions, once a luminosity function has been assumed, can be derived through the multi-messenger and multi-wavelength approach.  If the emission from extragalactic unidentified sources is primarily of hadronic origin, it will be accompanied by neutrinos at comparable fluxes, and that may have potentially observable consequences in the TeV range for future km$^3$ neutrino detectors such as IceCube and Km3Net. If on the other hand the gamma-ray emission is primarily leptonic, it will be accompanied by X-ray emission, and their extragalactic background flux in the X-ray band may provide an additional constraint.
We will pursue such models and calculations in an upcoming publication.  

In this work we have tried, where possible, to make 
assumptions, which, if anything, {\em underestimate} the possible
contribution of unresolved unidentified sources to the EGRB. An exception to this general trend is our 
 working assumption that 
the majority of the resolved, high-latitude 3EG unidentified sources belong
to a single class. It is conceivable that instead, the resolved
unidentified sources are a collection of members of several known and
unknown classes of gamma-ray emitters. In this case, it is
still likely that the summed contribution of unresolved members of all
parent classes to the diffuse background is significant. However, the
construction of a single cumulative flux distribution from all sources 
and its extrapolation to lower fluxes is no longer an indicative test
for the importance of such a contribution. 

Although we have argued that the contribution of unresolved
unidentified sources to the EGRB is likely to be important or even
dominant, the cumulative emission spectrum 
derived here and presented in Fig.\ \ref{ress} is only an upper limit,
as it was derived demanding that the observed EGRB is not exceeded at
any energy above 100 MeV. The contribution of unresolved unidentified
sources is further constrained by allowing for the presence of the
guaranteed contributions of unresolved normal galaxies and blazars. 

Finally, it is noteworthy that we have found no evidence for an
inconsistency between the population properties of high-latitude
unidentified sources and those of blazars. 
The presence of a yet-unknown population of extragalactic high-energy
emitters among the high-latitude EGRET unidentified sources remains
one of the most tantalizing possibilities in GeV astronomy and one of
the most exciting prospects for the GLAST era. However, 
the presently available data on the spectral index
distribution and the cumulative flux distribution of these sources 
(this work), as well as their variability properties (e.g., Nolan et
al.\ 2003), are consistent with their being members of the blazar population. If indeed a large number of members of the blazar class are present among the entragalactic unidentified sources, then this could have potentially serious effects on our understanding of the redshift distribution of resolved blazars and consequently of the blazar luminosity function.

\begin{acknowledgements}
We thank Chuck Dermer, Brenda Dingus, Stefano Gabici, Demos Kazanas, 
Tijana Prodanovi\'{c}, Olaf Reimer, Kostas Tassis, Tonia Venters, and an anonymous referee for helpful comments and 
discussions. This work was supported in part by the Kavli Institute for Cosmological Physics through the grant NSF PHY-0114422  and NSF PHY-0457069. 
\end{acknowledgements}

\begin{appendix}

\section{Estimating the spectral index distribution of extragalactic unidentified sources}\label{spindex}

\begin{figure}
\begin{center}
\resizebox{3.0in}{!}
{
\plotone{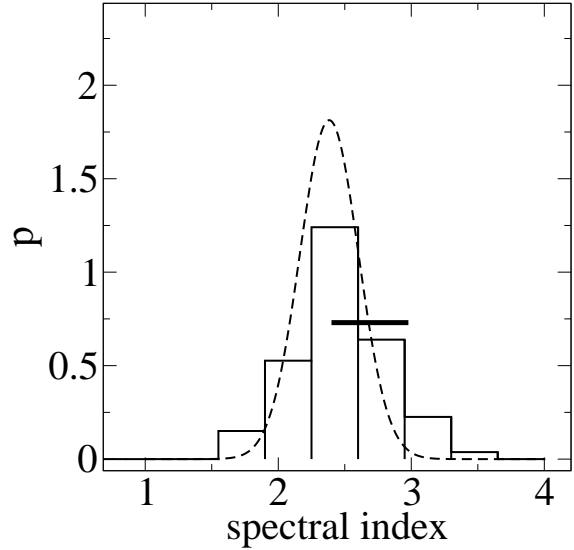}}
  \caption{\label{hist}
Spectral index distribution of resolved unidentified sources in Sample
1 (all sources unidentified in 3EG and all potential blazar associations). Solid
line: histogram of the EGRET data. Dashed line: maximum-likelihood
Gaussian. Thick solid line: typical uncertainty of individual spectral index
determination using EGRET data.}
\end{center}
\end{figure}

Figure \ref{hist} shows 
a histogram of the spectral indices of the
sources in Sample 1 (solid line). The typical 
measurement uncertainty for any single spectral index (thick solid line)
is  comparable with the spread of the
distribution, so the spread of a simple binning of spectral 
indices might not
in fact give us information about the underlying distribution of the
spectral indices of the sources, but rather be representative
of the uncertainty of each single measurement.  

Following an analysis similar to that of Venters \& Pavlidou (2007)
for the case of blazars, we assume that the
intrinsic spectral index distribution (ISID) of unidentified sources can be
approximated by a Gaussian,
\begin{equation}\label{spindexes}
p(\alpha) = \frac{1}{\sqrt{ 2 \pi} \sigma_0}
\exp\left[-\frac{(\alpha - \alpha_0)^2}{2 \sigma_0^2}\right]\,.
\end{equation}
Lacking any knowledge to the contrary, this choice not only simplifies the subsequent statistical analysis, but also renders the spectral shape part of Eq.\ (\ref{contr}) analytically integrable (see Appendix).
We then use a  maximum-likelihood  analysis which takes 
into account the individual errors of 
measurement of $\alpha$ for each source by introducing the true spectral 
indices of the sources as nuisance parameters and by marginalizing 
over them. The likelihood in this case, omitting constant
normalization factors, is  (Venters \& Pavlidou 2007)
\begin{equation}\label{likel}
\mathcal{L} 
= 
\left(
\prod_{j=1}^{N}\frac{1}{\sqrt{\sigma_0^2+\sigma_{j}^2}}
\right)
\exp\left[-\frac{1}{2} \sum_{j=1}^{N}
\frac{(\alpha_{j} - \alpha_{0})^2 }{\sigma_0^2+\sigma_{j}^2}
\right],
\end{equation}
where $\alpha_j$ and $\sigma_j$ are the measured spectral index and the
associated uncertainty of this measurement for a single source. 
Table \ref{isids} shows the maximum-likelihood $\alpha_0$ and
$\sigma_0$ for each sample. 
The maximum-likelihood spectral index distribution for sample 1 is
plotted with the dashed line in Fig.\ \ref{hist}, and  
indeed is narrower than 
the histogram. 

\begin{figure}
\begin{center}
\resizebox{3.3in}{!}
{
\plotone{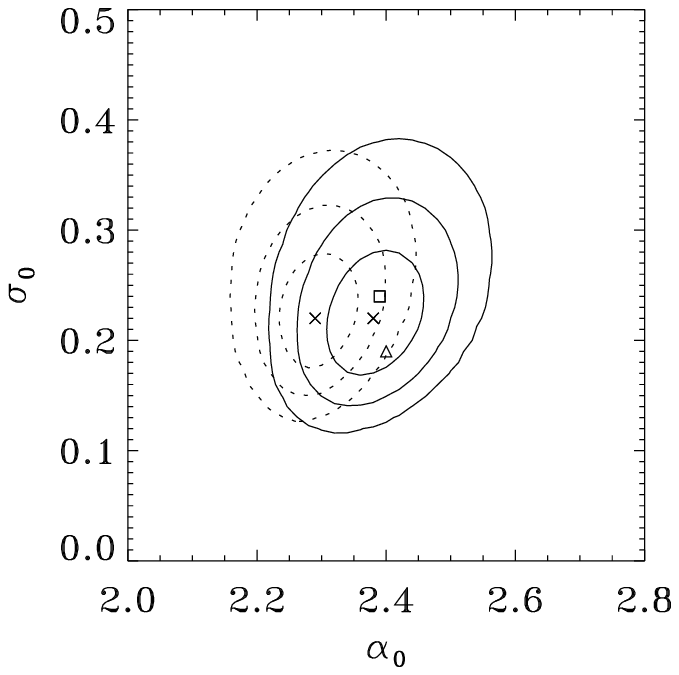}
}
  \caption{\label{cont}
Likelihood contours for the intrinsic spectral index distribution
parameters of Sample 1 (solid line). The maximum-likelihood parameters
are denoted by $\times$. The $\square$ and $\triangle$ symbols denote
the locations of the maximum-likelihood parameters for Sample 2 and
Sample 3 respectively. As a comparison, the likelihood contours for
the sample of confidently identified blazars (as in Mattox et al.\ 2001) are
overplotted with the dashed lines. }
\end{center}
\end{figure}

Figure \ref{cont} shows $1\sigma$, $2\sigma$, and $3\sigma$ contours
for the likelihood of Eq.\ \ref{likel}, as calculated using Sample 1
data (solid line). The location of the maximum is indicated by
$\times$, while the corresponding maxima of the Sample 2 and Sample 3
likelihoods are indicated by $\square$ and $\triangle$
respectively. In both cases, the location of the maximum is within the
$1\sigma$ contour of Sample 1. As an additional comparison, we
overplot, with the dashed lines, the $1$, $2$, and $3\sigma$
likelihood contours for the sample of the 46 confidently identified blazars of
Mattox et al.\ (2001). The figure demonstrates that 
extragalactic unidentified sources are spectrally consistent, at the
$1\sigma$ level, with being blazars. 

\begin{table}
\begin{center}
\caption{Maximum-likelihood parameters for the intrinsic spectral
  index distribution \label{isids}}
\begin{tabular}{|l|ccc|}
\hline
& $\,\,\,\,\,$ Sample 1 
$\,\,\,\,\,$
& $\,\,\,\,\,$ Sample 2  $\,\,\,\,\,$& 
 $\,\,\,\,\,$Sample 3 $\,\,\,\,\,$\\
\hline
$\alpha_0$ & $2.38$ & $2.39$ & $2.40$ \\
$\sigma_0$ & $0.22$ & $0.24$ & $0.19$ \\
\hline 
\end{tabular}
\end{center}
\end{table}

\section{Source Variability}\label{Var}

Certain classes of sources contributing to the gamma-ray background
may exhibit significant time variability (blazars are a prominent
example of such sources). This fact has generated some concern that 
the objects resolved by EGRET (especially the fainter ones) are 
preferentially objects that happened
to be in their flaring state when observed, and hence the luminosity 
function as determined by the 
resolved objects is not an adequate description of all the unresolved
sources contributing to the gamma-ray background. Stecker \& Salamon (1996a) have addressed that uncertainty in the case of blazars using an explicit treatment of source variability. 

This concern is less pronounced if the fluxes used for 
normalizing the luminosity function of a class of sources are 
averaged over exposure times 
reasonably large compared to the source duty cycle. 
To minimize the possible impact of variability, we use in this study
the P1234 fluxes quoted in 3EG. Sources which have been detected in  
individual viewing periods but which have no P1234 flux (which implies
 that their detection significance in the cumulative map was not high
 enough) are omitted from our analysis. 

We have not made any further attempt to treat variability in this
work, because the unknown parameters in the variability
properties of sources are severe enough
that any simple treatment of variability is likely to introduce at
least as much uncertainty as it attempts to remove, while any
uncertainty in the flux distribution of the source samples  due to source
variability is minor compared to the uncertainties
introduced due to our ignorance concerning the nature of unidentified 
sources. We have simply used the time-averaged flux of resolved 
unidentified sources as representative of the emission properties
 of such sources. 

\section{Spectral shape of cumulative unresolved point source emission}\label{shape}

The spectral shape of the cumulative emission from a population of 
unresolved sources with individual power-law photon 
spectra $F_E \propto E^{-\alpha}$ and a Gaussian spectra index
distribution $p(\alpha)$ is given by Eq.\ (\ref{contr}).
If $p(\alpha)$ is a Gaussian of the form of
Eq.\ (\ref{spindexes}), as assumed here, then Eq.\ (\ref{contr}) can be integrated
analytically. Defining 
$\tilde{I}_0 = I_0 \left[(\kappa CF_8^{-\kappa+1})/(1-\kappa)\right]_{F_{8, \rm
    min}}^{F_{8, \rm max}}$, we can write
\begin{eqnarray}\label{apeq}
I_E(E) &=& \tilde{I}_0\int_{-\infty}^{\infty} d\alpha (\alpha-1) 
\frac{1}{\sqrt{2\pi}\sigma_0}\exp\left[-\frac{(\alpha-\alpha_0)^2}{2\sigma_0^2}\right]\exp\left[-\alpha\ln
  \frac{E}{E_0}\right] \nonumber \\
&=& \frac{\tilde{I_0}}{\sqrt{2\pi}\sigma_0} 
\exp\left[-\frac{B_0}{2\sigma_0^2}\right]
\int_{-\infty}^{\infty} d\alpha (\alpha-1)
\exp\left[-\frac{(\alpha-A_0)^2}{2\sigma_0^2}\right] 
= \tilde{I}_0(A_0-1)\exp[-\frac{B_0}{2\sigma_0^2}]\,,
\end{eqnarray}
where $A_0=\alpha_0-\sigma_0^2\ln (E/E_0)$ and $B_0 =
2\alpha_0\sigma_0^2\ln(E/E_0) - \sigma_0^4[\ln(E/E_0)]^2$. The usual
way of plotting the isotropic background spectrum is $E^2I_E$ vs $E$
in logarithmic scale (as in Fig.\  \ref{ress}). Defining
$X = \ln(E/E_0)$ and $Y=\ln(E^2I_E)$, we can rewrite Eq.\ (\ref{apeq}) as
\begin{equation}\label{anres}
Y = \frac{\sigma_0^2}{2}X^2 - (\alpha_0-2)X+C
+\ln(\alpha_0-1-\sigma_0^2X)\,, 
\end{equation}
where $C = \ln(\tilde{I}_0E_0^2)$. Note that the limits of integration of Eq.\ \ref{apeq} should strictly extend from $\alpha=1$ (instead of $-\infty$) to $\infty$, to ensure that all $F(>100 {\, \rm MeV})$ are positive and finite. The result of Eq.\ (\ref{apeq}) is a good approximation as long as $\alpha_0$ is high enough and $\sigma_0$ is low enough so that $p(\alpha)$ has no appreciable power at $\alpha < 1$. Equation (\ref{anres}) consists of
a parabolic convex part and a logarithmic concave part. The parabolic
part dominates at low energies and is
responsible for the characteristic convex shape of the unresolved
emission from a population of sources with power-law spectra and a
finite-spread spectral index distribution (see, e.g., Brecher \&
Burbidge 1972; Stecker \& Salamon 1996 a,b; Pohl et al.\ 1997).
 The spectrum remains convex ($d^2Y/dX^2$ is positive) for
$X<(\alpha_0-1)/\sigma_0^2 -1$. The curvature of the spectrum depends
on the spread of the spectral index distribution, $\sigma_0$ (the
second-order term coefficient in Eq.\ \ref{anres} is
$\sigma_0^2/2$). Consequently, an overestimation of the spread of the
population spectral index distribution will lead to an overestimation
of the curvature of the corresponding unresolved emission spectrum (as
pointed out by Pohl et al.\ 1997, Pavlidou et al.\ 2006, and 
Venters \& Pavlidou 2007). In the
limit $\sigma_0 \rightarrow 0$ all sources have identical
spectra of index $\alpha_0$, and Eq.\ (\ref{anres}) yields a power-law, 
$Y={\rm const}
+ (\alpha_0-2)X$ or, equivalently,  $I_E \propto E^{-\alpha_0}$, as expected. 
For $\sigma < 0.5$, 
Eq.\ (\ref{anres}) has a local minimum, at $X = (2\alpha_0-3
- \sqrt{1 -4 \sigma_0^2})/(2\sigma_0^2)$,
which corresponds to the energy at which the population has the
smallest energy flux contribution to the (unabsorbed) isotropic
background. 
\end{appendix}

\end{document}